# One-for-multiple substitution in solid solutions

W. T. Geng[*] and Q. Zhan

*Materials Science and Engineering, University of Science and Technology Beijing, Beijing 100083, China*

**Abstract**

It is generally assumed that one solute atom will occupy only one lattice site in a substitutional solid solution. We here report an interesting discovery by first-principles calculations that a large solute atom can replace multiple matrix atoms in the elemental crystal of beryllium. Examination on Groups IIIB, IVB, VB, VIB, and VA elements shows that Cr will substitute for one, V and Mo for three, Sc, Y, Ti, Zr, Hf, W, Nb, Ta, As, Sb, and Bi for four, and La for five Be atoms. Dissolution of Zr, Hf, Sc, and Y is exothermic, suggesting a good solubility. At low concentration, the configurational entropy resulted from one-for-multiple substitution is larger than in the one-for-one substitution case. We find that Sc, Y, Zr, and Hf all have tendency to aggregate in Be, but Sc is the weakest among them and thus can be expected to improve the superplasticity of Be.

[*] geng@ustb.edu.cn

**I. Introduction**

By definition, a substitutional solid solution is a mixture of two types of atoms in which the solute atoms replace the solvent (matrix) ones in some lattice sites [1]. Although it has not been stated explicitly, that one substitutional solute atom will replace only one matrix atom is commonly accepted. If the solute atoms are too small compared with the matrix atoms, they will prefer to occupy interstitial positions, and hence the formation of an interstitial solid solution. But, what if the solute atom is too large, say, more than 15% larger in radius (50% larger in volume) than a matrix atom? Then, the solubility of this type of element will be extremely low according to the Hume-Rothery rules [2]. One might still be curious about the condition where the solute atom is extraordinarily larger than the matrix atom. Could there be a one-for-multiple replacement in such a substitutional solid solution?

The nearest-neighbor distance between matrix atoms, a fairly good measure of the atomic diameter [3], can be as small as 1.54 Å in diamond, and as large as 3.65 Å in yttrium or similarly in lanthanides [1]. Due to the high strength of carbon-carbon bonds, only boron, nitrogen, phosphorus, and sulfur can be effectively doped into diamond [4]. Since these elements are not exceedingly larger than carbon, there should be no chance for any of them to substitute for more than one carbon atom upon doping. The other elemental crystal with a nearest-neighbor distance just slightly larger than diamond is the $\beta$ phase of boron, a rhombohedral lattice in which the shortest boron-boron bond is 1.63 Å [5]. However, due to similar reason as in the case of diamond, doping or alloying extraordinarily large elements into boron to modify its properties is unfeasible.

Then comes beryllium, the elemental crystal with the third smallest atomic diameter, 2.26 Å [1]. A combination of desirable properties such as high strength-to-weight and stiffness-to-weight ratios, high specific heat, and high thermal conductivity, making beryllium the material of choice for x-ray and nuclear applications. Beryllium form alloys and intermetallic compounds with a great many elements [6]. Among the allying elements added into beryllium, Sb [7] and Y [8] arouse our strong curiosity, for they have much larger atomic size than Be. According to the Hume-Rothery rules [2], they could only have very marginal solubility in Be. If the substitution of Sb and Y in Be is in a one-for-one manner, the elastic strain energy [9] caused by the huge size-mismatch will lead to very high formation heat in Be-Sb and Be-Y solid solutions. In the original disclosure [7], it was not clear whether the composition of Be-Sb was an alloy or a composite, or a mixture of both. But in the Be-Y case, at least a remarkable amount of Y atoms are in solution state [8]. We speculate that in Be, Sb and Y might substitute for not a single Be atom, but for a small Be clusters.

**II. Computational**

To verify this conjecture, we have performed first-principles density functional theory (DFT) calculations on the formation energy of Groups IIIB, IVB, VB, VIB, and VA elements (Sc, Y, La, Ti, Zr, Hf, V, Nb, Ta, Cr, Mo, W, As, Sb, and Bi) in Be and examined one-for-$x$ ($x$=1-5) substitution types. The computational technique employed in this work is DFT based Vienna Ab initio Simulation Package [10]. The electron-ion interaction was described using projector augmented wave (PAW) method [11]. The exchange correlation between electrons was treated with generalized gradient approximation (GGA) in the Perdew-Burke-Ernzerhof (PBE) form [12]. To calculate the solution energies of alloying elements, we have employed (5×5×3) supercells of hcp Be, an alloying atom in which represents a concentration of 0.67 at%. We used an energy cutoff of 300 eV for the plane wave basis set for pure beryllium and all solution systems to ensure equal footing. The Brillouin-zone integration was performed within the Monkhorst-Pack scheme [13] using *k* mesh (3×3×3) for these supercells. Since the size mismatch between alloying elements and Be is significant, both the shape and volume of the supercell, besides the internal coordinates of all atoms, are optimized. The energy relaxation for each supercell was continued until the forces on all the atoms were converged to less than $1\times10^{-2}$ eV Å$^{-1}$.

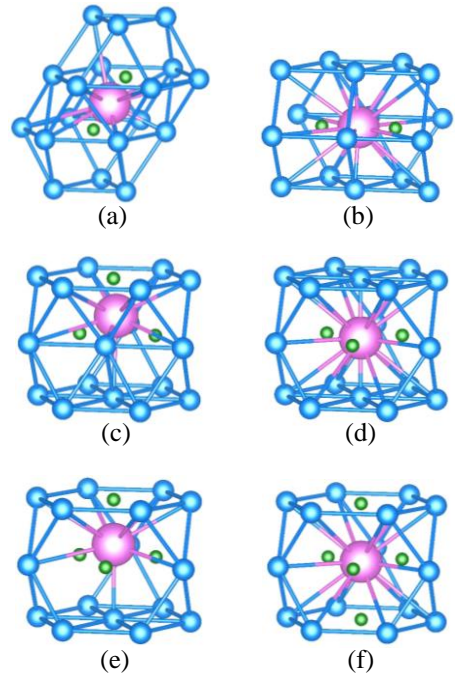

(a)  (b)

(c)  (d)

(e)  (f)

**Fig. 1. The local structure around the solute atom in the one-for-two (a and b), one-for-three (c and d), one-for-four (e) and one-for-five (f) substitutional types in a (5×5×3) supercell of hcp beryllium. The small, medium, and large circles represent substituted Be, matrix Be, and the solute atoms.**

When one solute atom replaces more than one Be atoms, those lattice sites should be as closely packed as possible in order to minimize the surface of this tiny cluster. We illustrate the one-for-*x* (*x*=2-5) substitution types in beryllium solid solution in Fig. 1. The matrix Be atoms are displayed as medium circles, those that will be replaced as small ones, and the solution atom as a large circle. The one-for-one substitution (not shown) is just the situation as illustrated in all textbooks. In the case of one-for-two replacement, the two Be atoms forming a dumbbell can be in two neighboring (0001) planes (panel a) or in the same (0001) plane (panel b). Likewise, in the one-for-three case, the three Be atom can be in two neighboring (0001) planes (panel c) or in the same (0001) plane (panel d). Since four Be atoms can make a tetrahedron in close-packing, we consider only such a configuration in which three Be are in one (0001) plane and the fourth atom in one of the two neighboring (0001) planes (panel e). Obviously, adding one more Be atom to a tetrahedron constructs a triangular bipyramid (panel f). The solution energy of *M* (*M*= Sc, Y, La, Ti, Zr, Hf, V, Nb, Ta, Cr, Mo, W, As, Sb, Bi) in a (5×5×3) supercell of hcp Be $\Delta H_f(M, x)$, in a one-for-*x* (*x*=1-5) type substitution, is defined as

$$\Delta H_f(M, x) = E(\text{Be}_{5\times5\times3} - x\text{Be} + M) - E(\text{Be}_{5\times5\times3}) + xE(\text{Be}) - E(M) \qquad (1)$$

where $E$ is the total free energy of appropriate supercells and that of Be and M in their elemental crystals. The calculated solution energies are listed in Table 1. We have also calculated the relative volume size mismatch, termed as volume size factor by King [14]. It is defined as

$$\Omega_{sf}(M, x) = (\Omega_M - x\Omega_{\text{Be}})/\Omega_{\text{Be}} \qquad (2)$$

in which $\Omega_M$ and $\Omega_{\text{Be}}$ are the atomic volume of elemental crystals *M* and Be.

### III. Results

**Table 1.** The calculated solution energy $\Delta H_f(M,x)$ (in eV) and volume mismatch $\Omega_{sf}(M,x)$ (in %) of Groups IIIB, IVB, VB, VIB, and VA elements in a (5×5×3) supercell of hcp Be, with one solute atom *M* replacing *x* (*x*=1-5) Be atoms. The minimum solution energy and volume mismatch for each solute are emphasized in bold fonts. Negative solution energies indicate exothermic processes.

| $x=$ | 1 | | 2 | | 3 | | 4 | | 5 | |
|---|---|---|---|---|---|---|---|---|---|---|
| M | $\Delta H_f$ | $\Omega_{sf}$ | $\Delta H_f$ | $\Omega_{sf}$ | $\Delta H_f$ | $\Omega_{sf}$ | $\Delta H_f$ | $\Omega_{sf}$ | $\Delta H_f$ | $\Omega_{sf}$ |
| Sc | 3.58 | 197.7 | 2.09 | 48.9 | 0.34 | **0.8** | **-0.26** | -25.6 | 0.66 | -40.5 |
| Y | 6.16 | 288.4 | 3.32 | 94.2 | 1.36 | 29.5 | **-0.10** | **-2.9** | 0.26 | -22.3 |
| La | 8.07 | 359.8 | 4.16 | 129.9 | 1.95 | 53.3 | 1.04 | +14.9 | **0.93** | **-8.0** |
| Ti | 2.46 | 116.2 | 1.58 | **8.1** | 0.40 | -27.9 | **0.25** | -46.0 | 0.86 | -56.8 |
| Zr | 4.52 | 195.1 | 2.20 | 47.5 | 0.37 | **-1.6** | **-0.58** | -26.2 | 0.20 | -41.0 |
| Hf | 4.30 | 194.2 | 2.24 | 47.1 | 0.51 | **-1.9** | **-0.24** | -26.5 | 0.41 | -41.2 |
| V | 2.04 | 65.6 | 1.86 | **-17.2** | **1.23** | -44.8 | 1.54 | -63.9 | 2.53 | -66.9 |
| Nb | 3.16 | 127.3 | 1.79 | **13.7** | 0.47 | -24.2 | **0.32** | -43.2 | 1.16 | -54.5 |
| Ta | 3.68 | 127.9 | 2.30 | **14.0** | 0.99 | -24.0 | **0.79** | -43.0 | 1.59 | -54.4 |
| Cr | **1.62** | 44.2 | 2.12 | **-27.9** | 2.26 | -51.9 | 2.67 | -63.9 | 3.61 | -71.2 |
| Mo | 2.39 | 96.3 | 2.16 | **-1.8** | **1.54** | -34.6 | 1.68 | -50.9 | 2.46 | -60.7 |
| W | 3.03 | 101.1 | 2.74 | **0.6** | 2.27 | -33.0 | **2.10** | -49.7 | 2.84 | -59.8 |
| As | 4.23 | 180.4 | 3.97 | 40.2 | 3.99 | **-6.5** | **2.41** | -29.9 | 3.70 | -43.9 |
| Sb | 6.57 | 301.3 | 5.49 | 100.6 | 4.79 | 33.8 | **2.88** | **0.3** | 2.51 | -19.7 |
| Bi | 8.19 | 360.3 | 6.28 | 130.2 | 5.59 | 53.4 | **3.53** | +15.1 | 3.81 | **-7.93** |

It turns out that our conjecture prior to the first-principles calculations is correct. Among these 15 species, only Cr, which has an atomic size not much larger than that of Be, will replace only one Be atom, V and Mo will substitute for three, La for five, and the other 11 elements will make one-for-four substitution in hcp Be. We note that for the one-for-two substitution, the configuration with a smaller Be-Be bondlength (Fig. 1a) is always slightly (several percent eV) more favorable than the one in Fig. 1b; whereas for the one-for-three substitution, the geometry in Fig. 1d is slightly preferred. Listed in Table 1 do not include the values for configurations in Fig. 1b and Fig. 1c. For each species, the minimum solution energy value and the smallest size-mismatch is emphasized in bold fonts. We see in Table 1 that Zr has the lowest solution energy (-0.58 eV/atom), followed by Sc (-0.26 eV), Hf (-0.24 eV), and Y (-0.10 eV). Antimony should have very limited solubility due to high energy cost in dissolution. Therefore, we conclude that the Be-Sb composition produced in [7] was essentially a composite, not an alloy.

It is seen in Table 1 that except for Y, La, and Sb, where the best size match comes together with the most energetically favorable substitution, other 12 solutes do not have such a concurrence. Nonetheless, eight of these 12 species, namely Sc, Zr, Hf, V, Cr, Mo, As, and Bi, have the second best size match in the favorable substitution. And even for the rest four solutes, Ti, Nb, Ta, and W, the volume mismatch is always within 50%. This means that the one-for-multiple substitution still abide the primary Hume-Rothery rule. We point out that a pair of nearest-neighbor Be atoms appear as a dumbbell, three Be as a triangle, four Be as a tetrahedron, and five Be as a triangular bipyramid. Among these four geometries, a tetrahedron has the best sphericity, and hence the smallest specific surface area. As a consequence, the one-for-four substitution is heavily favored and the majority of large solute atoms take this form.

In a usual solid solution with one-for-one substitution, the configurational entropy per atom (lattice site) is

$$S_{mix}^1 = -k[x \ln x + \ln(1-x) - x \ln(1-x)] \qquad (3)$$

where $x$ is the atomic concentration of the solute and $k$ is the Boltzmann constant. In the case of one-for-four substitution, it is not as simple. Suppose there are $m$ Sc atoms in a bulk of hcp Be with $N$ lattice sites, the atomic concentration of Sc will be

$$x = \frac{m}{N-3m} \qquad (4)$$

Since there are 2N tetrahedrons in the bulk of N lattice sites, the configurational entropy per lattice site is

$$S_{mix}^4 = -k[x' \ln x' + 2\ln(1 - x'/2) - x' \ln(2 - x')] \qquad (5)$$

where is $x'$ defined as

$$x' = \frac{m}{N} = \frac{x}{1+3x} \qquad (6)$$

At low concentration, $x' \approx x$, and

$$S_{mix}^4 - S_{mix}^1 \approx kx[\ln(2-x) - \ln(1-x)] > 0. \qquad (7)$$

This means that one-for-four substitution results a larger configurational entropy.

**IV. Discussion**

Since Zr, Hf, Sc, and Y have negative solution energies, we expect that these elements could have more remarkable solubility than other large atoms in Table 1. It is known that all of them can form intermetallic compounds with Be [6]. The strong tendency to form intermetallic compound, however, is a strong hindrance when the state of solid solution is desired. This is the situation when appropriate alloying additions are sought to mitigate the grain growth in Be in order to realize an optimal superplasticity [8]. The aggregation of solute atoms into the formation of intermetallic compound starts with the pairing of solute atoms. Thus, we here evaluate the paring energy (attraction) of two solute atoms in a (5×5×3) supercell of hcp Be.

$$E_p(M) = \Delta H_f(M,x) + \Delta H_f(M,y) - \Delta H_f(2M,x,y) \qquad (8)$$

where $\Delta H_f(2M,x,y)$ is the solution energy of two *M* atoms substituting respectively for $x$ and $y$ Be atoms, defined similarly as in Eq. (1)

$$\Delta H_f(2M,x,y) = E(\text{Be}_{5\times5\times3} - (x+y)\text{Be} + 2M)$$
$$- E(\text{Be}_{5\times5\times3}) + (x+y)E(\text{Be}) - 2E(M) \qquad (9)$$

Based on the knowledge obtained from Table 1, we here have considered only two combinations, namely, (3+4) and (4+4), for variation of (*x*+*y*). We display in Fig. 2 the local geometries in a (5×5×3) supercell of hcp Be when a pair of large solute atoms replace 7 (panel a) or 8 (panel b) Be atoms.

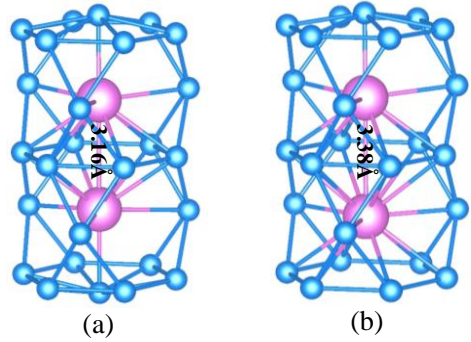

(a)        (b)

**Fig. 2. The local structure in the vicinity of a Sc-Sc pair substituting for 7 (panel a) and 8 (panel b) matrix atoms in a (5×5×3) supercell of hcp Be. The large and small circles represent Sc and Be atoms respectively. Shown here are already optimized structures.**

Our first-principles calculations demonstrate that for all these four species, the two-for-seven substitution is always more favorable than the two-for-eight type, while both of them are more favorable than two separate one-for-four substitutions. The paring energies calculated via Equations (3) and (4) are listed in Table 2. The driving force for solute paring is the formation of *M-M* bond, which is stronger than the Be-Be and *M*-Be bonds. The Sc-Sc bondlength is 3.16 Å in the two-for-seven substitution (Fig. 3a) and 3.38 Å in the two-for-eight substitution case (Fig. 3b). Our calculation shows that in elemental crystal of Sc, the nearest-neighbor distance is 3.17 Å, meaning that the Sc-Sc bondlength in the two-for-seven substitution case is very optimal, and hence a more favorable energetic state.

Clearly, Sc, Y, Zr, and Hf all have strong tendency to aggregate, and Sc has the weakest among them. This is in accordance with the fact that Sc has the lowest cohesive energy (weakest bonding) in its elemental crystal among four species. Given that Sc can be more easily dissolved into Be than Y and has a weaker tendency to aggregate at the same time, it is expected to have better performance in improve the high-temperature plasticity than Y. Zirconium may have similar effect to Sc, but the atomic weight makes it less ideal.

**Table 2. Pairing energy $E_p(M)$ (in eV) of solute atoms *M* in a (5×5×3) supercell of hcp Be. Positive values indicate attraction or binding.**

| *M* | Sc | Y | Zr | Hf |
|---|---|---|---|---|
| $E_p(M)$ | 0.60 | 1.11 | 0.89 | 1.06 |

**V. Concluding Remarks**

To summarize, our first-principles density functional theory calculations provide strong evidence that in Be solid solutions, a very large atom can substitute for a multiple of matrix atoms. This finding makes important supplement to the conventional wisdom that in a substitutional solid solution, a solute atom will replace one matrix atom. Such a knowledge is crucial to theoretical or computational studies on alloys and solid solutions in which one needs to formulate specifically the substitution type into the treatment.

The individual large solute atoms making one-for-multiple substitution in Be not only can modify the mechanical properties as alloying additions in a usual way, but also can behave as the tiniest precipitates. The substantial open volume around could enable them to serve as traps for H and He [15] under irradiation conditions. In addition, since the solution energies of one-for-three and one-for-five are not too different from that of the most favorable one-for-four substitution, these large atoms can also serve as strong traps and annihilation centers for self-interstitials and vacancies. In this sense, the knowledge that one large solute atom can substitute for several matrix atoms is imperative for the research on beryllium metallurgy, especially for atomic-scale computational studies on solute diffusion, precipitation, solute-defect binding, and also solute-solute interactions.